\documentstyle[aps,prl,multicol,epsf]{revtex}
\begin{document}
\title{\bf Self-organized Networks of Competing Boolean Agents}

\author{Maya Paczuski$^{1,2,3}$, Kevin E. Bassler$^3$, and \'Alvaro Corral$^1$ }
\address{$^1$ Niels Bohr Institute, Blegdamsvej 17, 2100
Copenhagen, Denmark \\
$^2$ NORDITA, Blegdamsvej 17, 2100 Copenhagen, Denmark \\
$^3$ Department of Physics, University of Houston, Houston TX
77204-5506 \\ }
\date{\today}

\maketitle %\parskip 2ex

\begin{abstract}
 
A model of Boolean agents competing in a market is presented where
each agent bases his action on information obtained from a small group
of other agents. The agents play a competitive game that rewards those
in the minority.  After a long time interval, the poorest player's
strategy is changed randomly, and the process is repeated.  Eventually
the network evolves to a stationary but intermittent state where
random mutation of the worst strategy can change the behavior of the
entire network, often causing a switch in the dynamics between attractors
of vastly different lengths.

\end{abstract}


%%%%%%%%%%%%%%%%%%%%%%%%%%%%%%%%%%%%%%%%%%%%%%%%%%%%%%%%%%%%%%%%
\begin{multicols}{2}

Dynamical systems with many elements under mutual regulation or
influence are thought to underlie much of the phenomena associated
with complexity.  Such systems arise naturally in biology, as, for
instance, genetic regulatory networks \cite{kauffman1}, or ecosystems,
and in the social sciences, in particular the economy \cite{arthur}.
Economic agents make decisions to buy or sell, adjust prices, and so
on based on individual strategies which take into account the
heterogeneous external information each agent has available at the
time, as well as internal preferences such as tolerance for risk.
External information may include both globally available signals that
represent aggregate behavior of many agents such as a market index, or
specific (local) information on what some other identified players are
doing.  In this case each agent has a specified set of inputs, which
are the actions of other agents, and a set of outputs, his own
actions, that may be conveyed to some other agents.  Thus, the economy
can be represented as a dynamical network of interconnected agents
sending signals to each other with possible, global feedback to the
agents coming from aggregate measures of their behavior plus any
exogenous forces.

 Each agent's current strategy can be represented as a function which
specifies a set of outputs for each possible input.  In the simplest
case the agents have only one binary choice such as either buying or
selling a stock \cite{bps}.  As indicated first by B.  Arthur this
simple case already presents a number of intriguing problems.  In his
``bar problem'', each agent must decide whether to attend a bar or
refrain based on the previous aggregate attendance history \cite{bar}.
Challet and Zhang made a perspicuous adaptation, the so-called
minority model, where agents in the minority are rewarded, and those
in the majority punished \cite{zhang}.  Common to all these and
related works \cite{savit} is that the network of interconnections
between the agents is totally ignored.  They are mean field
descriptions.  Each agent responds only to an aggregate signal,
e.g. which value (0 or 1) was in the majority for the last $T_i$ time
steps, rather than any detailed information he may have about other
specified agents.  It is not unexpected that an extended system with
globally shared information can organize.  A basic question in studies
of complexity is how large systems with only local information
available to the agents may become complex through a self-organized
dynamical process.

Here we explicitly consider the network of interconnections between
 agents, and for simplicity exclude all other effects.  We represent
 agents in a market as a random network of interconnected Boolean
 elements under mutual influence, the so-called Kauffman network
 \cite{kauffman1,kauffman2}.  The market payoff takes the form of a
 competitive game.  The performance of the individual agents is
 measured by counting the number of times each agent is in the
 majority.  After a time scale, defining an epoch, the worst
 performer, who was in the majority most often, changes his strategy.
 The Boolean function of that agent is replaced with a new Boolean
 function chosen at random, and the process is repeated indefinitely.
 Note that it is not otherwise indicated to the agents what is
 rewarded, i.e. being in the minority.  The agents are only given
 their individual scores and otherwise play blindly; 
 they do not know directly that they are rewarded by the outcome of 
 a minority game, unlike the original minority game model.

We observe that irrespective of initial conditions, the network
ultimately self-organizes into an intermittent steady state at a
borderline between two dynamical phases.  This border may correspond
to an ``edge of chaos'' \cite{kauffman1}.  In some epochs the
dynamics of the network takes place on a very long attractor;
 while, otherwise, the network is
either completely frozen or the dynamics is localized on some
attractor with a smaller period.  More precisely, numerical simulation
results indicate that the distribution of attractor lengths in the
self-organized state is broad, with no apparent cutoff other than the
one that must be numerically imposed, and consistent with power-law
behavior for large enough attractor lengths. A single agent's change
of strategy from one epoch to the next can cause the entire network to
flip between attractors of vastly different lengths.  Thus the network
can act as a switch.

Consider a  network of $N$ agents where each agent is assigned a
Boolean variable $\sigma_i=0 {\rm\ or\ } 1$.  Each agent receives
input from $K$ other distinct agents chosen at random in the system.
The set of inputs for each agent $i$ is quenched.  
The evolution of the system is specified by $N$ Boolean functions of
$K$ variables, each of the form
\begin{equation}
\sigma_i(t+1)= 
f_i\biggl(\sigma_{i_1}(t),\sigma_{i_2}(t),\cdots \sigma_{i_K}(t)\biggr)
\quad .
\end{equation}
There exists $2^{2^K}$ possible Boolean functions of $K$ variables.
Each function is a lookup table which specifies the
binary output for a given set of binary inputs.  In the simplest case
defined by Kauffman, {\it where the networks do not organize}, each
function $f_i$ is chosen randomly among these $2^{2^K}$ possible
functions with no bias; we refer to this case as the random Kauffman
network (RKN).

We will now briefly review some facts about Kauffman networks. First,
a phase transition occurs on increasing $K$.
For $K< 2$ RKN starting from random initial conditions reach frozen
configurations, while for $K> 2$ RKN reach attractors whose length
typically grow exponentially with $N$ and are called chaotic. RKN with
$K=2$ are critical and the distribution of attractors lengths that the
system reaches, starting from random initial conditions,  approaches
a power law \cite{powerlaw}, for large enough system sizes, when
averaged over many network realizations.  This phase transition in the
Kauffman networks can also be observed by biasing the random functions
$f_i$ so that the output variables switch more or less frequently if
the input variables are changed.  Boolean functions can be
characterized by a ``homogeneity parameter'' $P$ which represents the
fraction of 1's or 0's in the output, whichever is the majority for
that function.  In general, on increasing $P$ at fixed $K$, a phase
transition is observed from chaotic to frozen behavior.  For $K<2$ the
unbiased, random value happens to fall above the transition in the
frozen phase, while for $K\geq 3$ the opposite occurs
\cite{kauffman1}.    Kauffman networks are
examples of strongly disordered systems and have attracted attention
from physicists over the years (see for example
Refs. \cite{derrida,parisi,sneppen}).  Note that the phase transition
previously observed in Kauffman networks arises by externally
tuning parameters such as $P$ or $K$.

We consider  random Boolean networks of $K$ inputs, and with
lookup tables chosen independently from the $2^{2^K}$ possibilities
with equal
probability.  With specified initial conditions, generally random,
each agent is updated in parallel according to Eq. 1.  The agents are
competing against each other and at each time step those in the
minority win.  Thus there is a penalty for being in the herd.  One may
ascribe to agents a reluctance to change strategies.  Only in the face
of long-term failure will an agent overcome his barrier to change.  In
the limiting case of high barriers to change, the time scale for
changing strategies will be set by the poorest performer in the
network.  The change of strategies is approximated as an extremal
process \cite{bs} where the agent who was in the majority the most often over a
long time scale, the epoch, is chosen for ``Darwinian'' selection.  In
our simulations, the network was updated until either the attractor of
the dynamics was found, or the length of the attractor was found to be
larger than some limiting value which was typically set at 
10,000 time steps, solely for reasons of numerical convenience.
 The performance of the agents was then
measured over either the attractor or the portion of the attractor up
to the cutoff length.

The Boolean function of the worst player is replaced with a new
Boolean function chosen completely at random with equal probability
from the $2^{2^K}$ possible Boolean functions.  If two or more agents 
are the worst performers, one of them is chosen at random and changed.  
The performance of all the agents is then measured in the new epoch, and this
process is continued indefinitely.  Note that 
the connection matrix of the network does not evolve;
the set of agents who are inputs to each agent is fixed by the
initial conditions.  

\begin{figure}
\narrowtext \epsfxsize=2.65truein
\hskip 0.15truein \epsffile{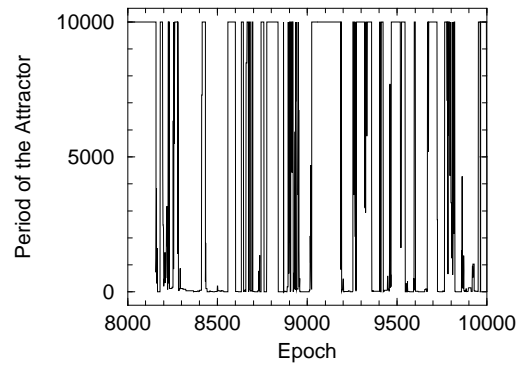}
\caption{ Time series of the length of attractor in each epoch for $K=3$,
$N=999$ in the stationary state.  }
\end{figure}

Independent of initial conditions, a $K=3$ network evolves to a
statistically stationary but intermittent state, shown in Fig. 1.
Initially the attractors that the system reaches are always very long,
consistent with all previous work on Kauffman networks.  But after
many epochs of selecting the worst strategy, short attractors
first appear and a new statistically
stationary state  emerges.  In this Figure we roughly
characterize an attractor as ``chaotic'' or long if its length is greater than
$l=10,000$ time steps.  On varying $l$ a similar
picture is obtained as long as $l$ is sufficiently large to
distinguish long period attractors from short period ones.
In the stationary state, one observes that the network can
switch behaviors on changing a single strategy.  Intriguingly,
Kauffman initially proposed random Boolean networks as simplified
models of genetic regulation where it is known that switches exist and
are important aspect of genetic control \cite{switch}.

 To be more precise, the histogram of the distribution of the lengths
of the attractor in the self-organized state was measured as shown in
Fig. 2 for different system sizes with the same numerically imposed
cutoff $l$.  The apparent peak at small periods is due to the relative
presence or absence of prime numbers, and numbers which can be
factored many ways.  The last point represents all attractors larger
than our numerically imposed cutoff 10,000, which is why a bump
appears.  In between these two regions, the behavior suggests a
power-law, $P_{atr}(t) \sim 1/t$ asymptotically, as is the case at the
phase transition in RKN \cite{powerlaw}.  If we increase or decrease
our numerically imposed cutoff then the bump at $l$ correspondingly
moves left or right and the intermediate region expands or contracts,
both consistent with the power law. Also the power law behavior becomes
more apparent for increasing system size suggesting that the
self-organized state we observed is not merely an effect of finite
system size, but becomes more distinct as the system size increases.
 
\begin{figure}
\narrowtext \epsfxsize=2.65truein
\hskip 0.15truein \epsffile{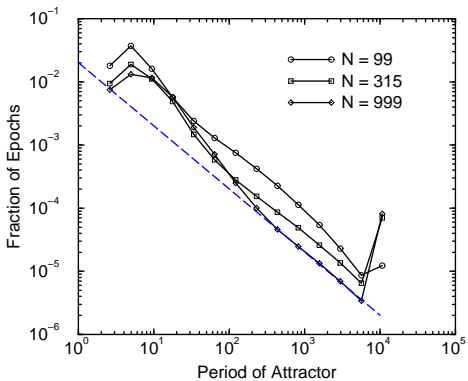}
\caption{ Histogram of Attractor Lengths for $K = 3$ Networks. 
The dashed line has a slope of 1.}
\end{figure}

The process of evolution towards the steady-state is monitored by
measuring the average value of the homogeneity parameter $P$ in the
network from epoch to epoch.  As shown in Fig. 3, for $K=3$, the
average value of $P$ tends to increase from the random value set by
the initial conditions during the transient.  For finite $N$, there
are fluctuations in $P$ in the steady state, as well as finite size
effects in the average value $\langle P\rangle$.  For
$N=(99,315,999,3161)$ we measured an average value in the steady state
$\langle P \rangle = (0.656(1), 0.664(1), 0.669(1), 0.671(1))$ and
root-mean-square fluctuations $\Delta P_{rms}\simeq (0.015, 0.007,
0.004, 0.001)$.  These numerical results suggest that in the
thermodynamic limit, $N\rightarrow\infty$, $P$ is approaching a unique
value $P_c \simeq 0.672$.  This value is below the $P_c \simeq
0.792438$ \cite{derrida2,note}
of random Kauffman $K=3$ networks, but is many standard
deviations away from the initial value.  

The dynamical state that the system evolves toward is different
from the phase transition  of Kauffman networks in other
(less trivial) ways.  In particular, the phase transition in RKN
 is a freezing transition where most elements do not change state.
  Only a few elements, strictly
$(<{\cal{O}}(N))$, are changing state \cite{kauffman1,derrida2}
at the phase transition of RKN, whereas
in our self-organized networks, there can be short attractors
associated with many elements $(\sim {\cal{O}}(N))$ changing state.
This can only occur if the Boolean tables in the network become
correlated by the evolutionary process, which, by construction, is not
allowed for RKN.  Thus our initially chaotic networks are not freezing
as in Kauffman networks at the phase transition, but are somehow phase locking
many elements together.

\begin{figure}
\narrowtext \epsfxsize=2.65truein
\hskip 0.15truein \epsffile{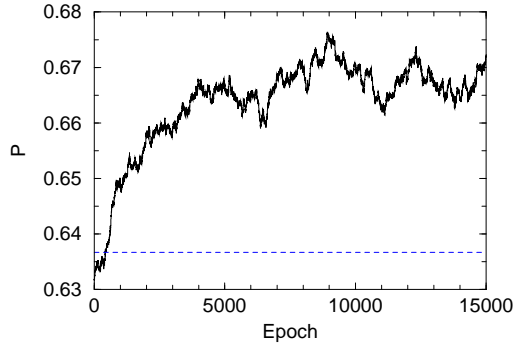}
\caption{ Self-organization of the homogeneity parameter $P$ for
same network as in Fig. 1.  The dashed line corresponds to the unbiased
random value.}
\end{figure}

The distribution of performances of agents in the network fluctuates a
great deal from epoch to epoch.  The performance is measured by
counting the fraction of times each agent is in the majority.  In the
case where the network has period one, there are obviously two peaks, one
corresponding to the group always in the minority and the other
corresponding to the group always in the majority. In fact we find
that even on the long attractors encountered in the steady state,
typically a significant fraction of the agents are frozen.  The number
of these frozen agents fluctuates from epoch to epoch.

 Fig. 4 is a histogram of performances for agents in a self-organized
network in a particular epoch which had a period greater than
$10,000$.  Note that the relative performances vary considerably.  The
two peaks represent the frozen agents.  As indicated in the figure,
the frozen agents are typically divided between the two states
unevenly.  In any given instant, despite the uneven division between
the frozen agents, the total number of agents in the two states (0,1)
is almost evenly divided with fluctuations that are much smaller than
in RKN.  Active agents, who are changing their state in response to
the inputs of others, comprise the remainder of the histogram outside
of the two peaks.  As shown in this figure, some agents who are
inflexible and do not respond to their environment perform better than
some agents who respond to their changing inputs and change states.
This suggests that somehow the losers are being exploited by some
information travelling in the network that they respond to.  Also,
somewhat counterintuitively, a large group of agents who take the same
action, corresponding to the left hand peak, can compete very well in
spite of the fact that the minority game tends to punish herd
behavior.

 \begin{figure}
\narrowtext \epsfxsize=2.65truein
\hskip 0.15truein \epsffile{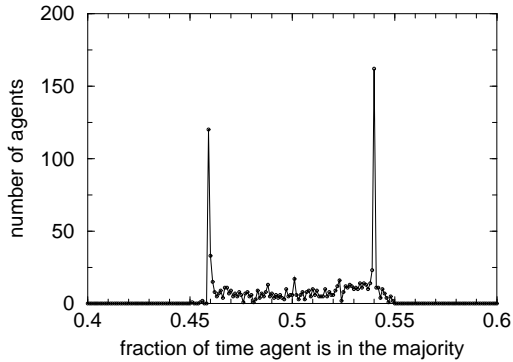}
\caption{ Histogram of performances in a particular epoch, for $N=999$ and
$K=3$ in the self-organized state. Those
with high scores are poor performers.}
\end{figure}

Although we currently have no adequate theoretical description of our numerical
observations, we can still discuss, to some extent, the
generality and robustness of our results.  First, if instead of
changing the entire Boolean table of the worst performer just one
element in it is changed, the self-organization process still takes
place.  If on the other hand, the Boolean function of the worst
performer and those who receive input from it are changed, no
self-organization takes place.  Of course it doesn't make sense to
change the Boolean functions of the agents who listen to the worst
performer because in our context the barrier to change is an internal
function of the performance for each individual.  The precise behavior
on varying $K$ is not determined at present.  For $K=6$, we have simulated
systems with $N=99$ as long as $10^6$ epochs and never observed the
system to reach any frozen state when starting from a random, unbiased
state in the chaotic phase, so it is possible that
the self-organization process as described here
using completely random tables does
not occur for high enough $K$.

However, other significant modifications were done where the
self-organization process survives.  For example, if instead of
changing the boolean tables of the worst performer, we keep the
boolean tables fixed at their initial state, 
 but change the inputs for the worst performer by
rewiring the network, then the $K=3$ networks still self-organize to a
similar state at an  ``edge of chaos'' with similar statistical properties
for the periods of the attractors and performances of the agents.
This occurs despite the fact that in this case the average homogeneity
parameter, $P$, of the network cannot evolve.

Rather than define an arbitrary fitness, and select those agents with
lowest fitness, an approach that was used by Bak and Sneppen
\cite{bs}, to describe co-evolution, we eliminate the concept of
fitness and define a performance based on a specific game.  Clearly if
the agents are rewarded for being in the majority then the behavior of
the system is completely trivial; the agents gain by cooperating
instead of competing and the network is driven deep into the frozen
phase.  This naturally raises the question of which types of games
lead to self-organized complex states.  In our model,
selection of agents in the majority for random change tends to
increase the number in the minority.  Even in the absence of
interactions, eventually those in the minority would become the
majority and lose.  We suspect that, in general, the game must make
agents compete for a reward that depends on the behavior of other
agents in a manner that intrinsically frustrates any group of agents
from permanently taking over and winning.  This frustration may be an
essential feature of the dynamics of many complex systems, and our
model may be interpreted, as, for instance, describing an
ecosystem of interacting and competing species.
  
We thank P. Bak, S. Kauffman, and K. Sneppen for stimulating
discussions. This work was funded in part by EU Grant No. FMRX-CT98-0183.

\end{multicols}

\begin{references}
\bibitem{kauffman1} S. A. Kauffman, {\it The Origins of Order,}
(Oxford University Press, New York, 1993).
\bibitem{arthur} W. B. Arthur, Science {\bf 284}, 107 (1999).
\bibitem{bps}P. Bak, M. Paczuski, and M. Shubik,  Physica A
{\bf 246}, 430 (1997).
\bibitem{bar}W. B. Arthur, Am. Econ. Rev. {\bf 84}, 406 (1994).
\bibitem{zhang} D. Challet and Y.-C. Zhang, Physica A {\bf 246}, 407
(1997); Y. -C. Zhang; Europhys. News {\bf 29}, 51 (1998).
\bibitem{savit} See for example R. Savit, R. Manuca, and R. Riolo, 
Phys. Rev. Lett. {\bf 82}, 2203 (1999).
\bibitem{kauffman2} S. A. Kauffman, J. Theoret. Biol. {\bf 22}, 437 (1969);
S. A. Kauffman, Physica D {\bf 10}, 145 (1984).
\bibitem{powerlaw}
A. Bhattachariya and S. Liang, Phys. Rev. Lett. {\bf 77}, 1644 (1996);
U. Bastolla and G. Parisi, J. Theor. Biol. {\bf 187}, 117 (1997).
\bibitem{derrida2} B. Derrida, in {\it Fundamental Problems in Statistical
Mechanics VII}, edited by H. van Beijeren (North-Holland, Amsterdam, 1990).
\bibitem{note}Note that the homogeneity parameter $P$ is not equal
to $p$ in Ref. \cite{derrida2} but can be related to it.
\bibitem{derrida} B. Derrida and Y. Pomeau, Europhys. Lett. {\bf 1}, 45
(1986).

\bibitem{parisi} U. Bastolla and G. Parisi, Physica D {\bf 115}, 203 (1998).
\bibitem{sneppen} S. Bornholdt and K. Sneppen, Phys. Rev. Lett. 
{\bf 81}, 236 (1998).
\bibitem{bs}
P. Bak and K. Sneppen, Phys. Rev. Lett. {\bf 71}, 4083 (1993);
M. Paczuski, S. Maslov, and P. Bak, Phys. Rev. E {\bf 53}, 414 (1996).
 
\bibitem{switch} M. Ptashne, {\it A Genetic Switch,} 
(Cell Press, Cambridge, 1992).

\end{references}
\end{document}